\documentclass[aps,prb,twocolumn,superscriptaddress,showpacs]{revtex4-1}

\usepackage{graphicx}
\usepackage{bm}
\usepackage{color}
\usepackage{amssymb} 
\usepackage{amsmath}
\usepackage{verbatim}

\renewcommand\Re{\operatorname{Re}}
\renewcommand\Im{\operatorname{Im}}
\newcommand{\la}{\langle}
\newcommand{\ra}{\rangle}
\newcommand{\rt}{\right}
\newcommand{\lf}{\left}

\renewcommand{\k}{\mathbf k}
\newcommand{\s}{\text{s}}
\renewcommand{\i}{\text{in}}

\begin{document}
\title{Imaging interatomic electron current in crystals with ultrafast resonant x-ray scattering }
\author{Daria Popova-Gorelova}
\email[]{daria.gorelova@desy.de}
\affiliation{Center for Free-Electron Laser Science, DESY, Notkestrasse 85, D-22607 Hamburg, Germany}
\affiliation{The Hamburg Centre for Ultrafast Imaging, Luruper Chaussee 149, D-22761 Hamburg, Germany}
\author{Robin Santra}
\email[]{robin.santra@cfel.de}
\affiliation{Center for Free-Electron Laser Science, DESY, Notkestrasse 85, D-22607 Hamburg, Germany}
\affiliation{The Hamburg Centre for Ultrafast Imaging, Luruper Chaussee 149, D-22761 Hamburg, Germany}
\affiliation{Department of Physics, University of Hamburg, D-20355 Hamburg, Germany}
\date{\today}
\begin{abstract}
We demonstrate how the technique of ultrafast resonant x-ray scattering can be applied to imaging dynamics of electronic wave packets in crystals. We study scattering patterns from crystals with electron dynamics in valence bands taking into account that inelastic and elastic scattering events induced by a broad-band probe pulse cannot be separated through the spectroscopy of the scattered photon. As a result, scattering patterns are not determined by the structure factor at the time of measurement, but can encode the instantaneous electron current between scattering atoms. We provide examples of how the interatomic electron current in a periodic structure can be extracted from a single scattering pattern by considering valence electron hole motion in (KBr)$_{108}$ and Ge$_{83}$ clusters. 
\end{abstract}
\maketitle

\section{Introduction}
The ability to image electronic dynamics in real space and real time is essential for a thorough investigation and control of  various chemical and physical transformations of molecular structures and crystals. This task requires angstrom spatial resolution, which corresponds to interatomic distances in solids and molecules, and femto- and even sub-femtosecond temporal resolution, which is necessary to capture electron dynamics. X-ray free electron lasers, capable of producing pulses of hard x rays with angstrom wavelengths, are a promising tool to achieve these requirements  \cite{CorkumNature07, GaffneyScience07, ChapmanNature06, VrakkingNature12, LeoneNature14}. Femtosecond time resolution is already achieved at free-electron laser facilities \cite{EmmaNature10, McNeilNature10}, and high-intensity attosecond pulses are realizable using the techniques described in Refs.~\onlinecite{ZholentsPRL04, EmmaPRL04, SaldinPhysRevSTAB06, KumarAppSciences13, TanakaPRL13,PratPRL15}.

Recently, we introduced a method that employs ultrafast hard-x-ray pulses for imaging dynamics of nonstationary electron systems in both real space and real time \cite{PopovaGorelova15}. This technique, combining elastic and inelastic ultrafast resonant x-ray scattering (RXS), allows encoding the interatomic electron current in a single scattering pattern in addition to the usual structural information. This effect is due to inelastic processes that unavoidably contribute to a scattering pattern obtained by a broadband ultrafast probe pulse. In order to demonstrate the principle of our method, we provided an illustrative example of probing electron hole motion in a diatomic molecule \cite{PopovaGorelova15}.

In this paper, we describe how our technique can be applied to imaging electron dynamics in crystals. Electron motion in crystals determines various fascinating ultrafast phenomena relevant for technological applications. Especially, this applies to light-driven phenomena, where transformations of electronic, optical or magnetic properties of a crystal are triggered by the interaction of an ultrafast pump pulse with electrons. Such light-induced phenomena include, for example, insulator-to-metal transition in charge-ordered systems \cite{KawakamiPRL10}, light-driven electron current in dielectrics \cite{SchiffrinNature12, SchultzeNature12}, high-harmonic generation \cite{GhimireNature11, SchubertNature14} and extraordinary carrier multiplication in semiconductors \cite{HiroriNature11}, and many other intriguing phenomena. In order to control such phenomena, the ability to characterize electron dynamics is necessary. 

In this article, we demonstrate how ultrafast RXS can be applied to image coherent electron motion in crystals and discuss conditions that are necessary to extract instantaneous interatomic electron currents from a scattering pattern. Launching and observing coherent electron dynamics in bulk crystals, which is a process of interest in view of fundamental physics as well as potential applications, has recently become possible \cite{PolliNature07, KawakamiPRL10, KuehnPRL10, KuehnPRB10, SchubertNature14}. For instance, phase-stable high harmonic transients covering the terahertz-to-visible spectral domain can be induced and controlled in a bulk semiconductor by triggering dynamical Bloch oscillations with THz pulses, which is an important step towards terahertz-rate electronics \cite{SchubertNature14}. It has also been suggested that dynamical Bloch oscillations have an effect on optical-field-induced currents in a wide-gap dielectric \cite{FoldiNJPh13, SchiffrinNature12}. It has been shown that coherent electron oscillations accompany insulator-to-metal transitions in a charge ordered organic crystal induced by few-optical-cycle infrared pulses \cite{KawakamiPRL10} and insulator-metal dynamics of a magnetoresistive manganite induced by an ultrashort optical pulse \cite{PolliNature07}. 


As examples of coherent electron dynamics in crystals, we consider coherent electronic wave packets in the valence bands of KBr and Ge crystals triggered by a photoionizing pump pulse. Creation of coherent electronic wave packets in valence orbitals of atoms and complex molecules by a photoionizing pump pulse has been demonstrated in attosecond science \cite{SmirnovaNature09, GoulielmakisNature10, SansoneNature10, HaesslerNature10, TzallasNature11, CalegariScience14}. Ultrafast control over ionization dynamics in solids by attosecond light pulses is a rapidly emerging field in attosecond science in view of potential applications for ultrafast signal processing \cite{GhimireJPhB14, YakovlevChapter}. Although some theoretical \cite{WachterPRL14, ZhokhovPRL14} and experimental \cite{MitrofanovPRL11, GertsvolfJPhB13} progress has been made in understanding attosecond ionization of solids, no optimal procedure for launching coherent electronic wave packets in crystals by photoionization has yet been established. Here, we do not aim at describing the formation of electronic wave packets in solids through photoionization, but we use them just as examples to analyze which dynamical information is encoded in time-resolved RXS patterns from electronically nonstationary periodic structures.

In Section \ref{Section_RXS}, we discuss the formalism underlying ultrafast RXS from electronic wave packets and show the connection between the interatomic electron current and the Fourier transform of a scattering pattern from $\mathbf Q$ space to real space \cite{PopovaGorelova15}. We discuss in detail ultrafast RXS patterns and demonstrate how the interatomic electron current can be extracted from a scattering pattern using the example of coherent electron hole motion in KBr crystal in Sec.~\ref{Section_KBr}. Another example, which we provide in Sec.~\ref{Section_Ge}, is coherent electron hole motion in the two outermost valence bands of Ge crystal. We discuss the difference between the scattering patterns and the procedures to extract interatomic electron currents in KBr and Ge crystals. We simulate the regions where a single electron hole is distributed in the crystals by (KBr)$_{108}$ and Ge$_{83}$ clusters.

\section{Ultrafast resonant x-ray scattering}
\label{Section_RXS}
We consider the process of ultrafast RXS, where, first, a resonant x-ray probe pulse induces a transition of an electron from a core shell of a certain atomic species to a valence band, where electron dynamics is taking place. Then, the created core vacancy is filled by some electron accompanied by the emission of a photon that contributes to the scattering pattern. For a stationary measurement, the major contribution to the scattering pattern is due to elastic scattering events, since they sum up coherently. As a result, a stationary scattering pattern depends on the static structure factor, which is determined by the electron density. However, the situation is different for time-resolved RXS from a nonstationary electron system, the necessary condition for which is that the probe pulse duration must be much shorter than the characteristic time scales of electron dynamics. As a result, the bandwidth of the probe pulse is much larger than the maximum splitting of the energy levels involved in the dynamics, which leads to the indistinguishability of elastic and inelastic scattering processes. As a consequence, a scattering pattern obtained by an ultrashort resonant x-ray pulse does not encode a structure factor at the time of measurement and is not determined by the electron density at the time of measurement, in contrast to stationary RXS \cite{PopovaGorelova15}.

In Ref.~\onlinecite{PopovaGorelova15}, we investigated ultrafast RXS patterns obtained by a probe pulse from an electronic system excited by a pump pulse at time $t=0$. Let the many-body Hamiltonian of the system in the absence of an x-ray field be $\hat H_{\text{m}}$ with eigenstates $\Phi_I$ and eigenenergies $E_I$. Thus, the electronic system at time $t$ is described by a density matrix
\begin{equation}
\hat \rho^{\text{m}}(t) = \sum_{I,K} \mathcal I_{IK}(t) |\Phi_I\ra\la \Phi_K| \label{El_den},
\end{equation} 
where the elements $\mathcal I_{IK}(t)$ are determined by the pump pulse.  We demonstrated that the quantum electrodynamics treatment based on the density matrix formalism within second order time-dependent perturbation theory \cite{Mandel} is necessary to correctly describe scattering patterns obtained by an ultrashort x-ray pulse from this nonstationary electron system. The resulting differential scattering probability (DSP) for a probe pulse with intensity $I_\i(t) = I_0\,e^{-4\ln2 ((t-t_p)/\tau_p)^2}$, which arrives at time $t_p$ after the pump pulse, can be represented as \cite{PopovaGorelova15}
\begin{align}
&\frac{dP}{d\Omega}= \frac{\tau_p^2I_0}{4\ln2c^4}\sum_{C_q,C_r}\Bigl[\cos(\mathbf Q\cdot\{\mathbf R_{C_q}-\mathbf R_{C_r}\})\Re(\mathcal A_{qr}) 
\nonumber\\
&\qquad\qquad\qquad\qquad\quad-\sin(\mathbf Q\cdot\{\mathbf R_{C_q}-\mathbf R_{C_r}\})\Im(\mathcal A_{qr})\Bigr] ,\nonumber\\
&\mathcal A_{qr}(t_p) = H_{qr}\sum_{I,K}\mathcal I_{IK}(t_p)(\mathbf D_{KJ_{C_r}}\cdot\boldsymbol \epsilon_\i^*)(\mathbf D_{J_{C_q}I}\cdot\boldsymbol \epsilon_\i),\label{DSP_DipApp_extended}\\
&H_{qr} = \sum_{F,s_\s}(\mathbf D_{J_{C_r}F}\cdot\boldsymbol \epsilon_\s)(\mathbf D_{FJ_{C_q}}\cdot\boldsymbol \epsilon_\s^*)\nonumber\\
&\qquad\qquad\times\Delta\omega_{JF}^2\int_0^{\infty} \frac{d\omega_{\k_\s}\omega_{\k_\s}W(\omega_{\k_\s})e^{-\frac{\Omega_F^2\tau_p^2}{4\ln2}}}
{(\omega_{\k_\s}-\Delta\omega_{JF})^2+\Gamma_{J}^2/4},\nonumber
\end{align}
where $c$ is the speed of light, the sum over $C_q$ and $C_r$ goes over all scattering atoms in the system situated at positions $\mathbf R_{C_q}$ and $\mathbf R_{C_r}$, respectively (atomic units are used for this and following expressions). $J_{C_{q(r)}}$ are intermediate states with an electron hole in a core shell of atom $C_{q(r)}$, the energy splittings between which we assume to be much smaller than the bandwidth of the ultrashort probe pulse (thus, $E_F-E_{J_{C_q}} = E_F-E_{J_{C_r}}=\Delta \omega_{JF}$), and $\Gamma_{J_{C_q}}=\Gamma_{J_{C_r}}=\Gamma_{J}$ are the core-hole linewidths. $\mathbf D_{AB} = \la \Phi_A|\int d^3 r \hat\psi^\dagger\, \mathbf r \,\hat\psi|\Phi_B\ra$ designates the dipole matrix element between electronic states $|\Phi_A\ra$ and $|\Phi_B\ra$, where $\hat\psi$ ($\hat\psi^\dagger$) is the electron field annihilation (creation) operator. $\omega_{\k_\s}$ is the energy of a scattered photon with a wave vector $\k_\s$, $W(\omega_{\k_\s})$ represents the spectral acceptance range of the photon detector, $\Omega_F = \omega_{\k_\s}-\omega_\i+E_F-\la E\ra$, where $\omega_\i$ is the photon energy of the probe pulse and $\la E\ra$ is the mean energy of the electronic system described by Eq.~(\ref{DSP_DipApp_extended}). $\mathbf Q = \k_\i-\k_\s$, where $\k_\i$ is the wave vector of the probe pulse.

In contrast to the DSP from a stationary system, the DSP in Eq.~(\ref{DSP_DipApp_extended}) is not centrosymmetric and does not encode a structure factor at the time of measurement, {\it i.e.}~${dP}/{d\Omega}\not\propto |\sum_C f_C(t_p)e^{i\mathbf Q\cdot\mathbf R_C}|$, where $f_C(t_p)$ is the scattering amplitude of atom $C$ at the time of measurement $t_p$. However, some useful information can indeed be extracted from ultrafast RXS patterns. Namely, not only do they contain structural information, but they also can provide the electron current between scattering atoms at the time of measurement. In this paper, we provide examples of ultrafast RXS scattering patterns from crystals and discuss in detail what is actually needed to extract the interatomic electron current from a scattering pattern.

\subsection{Interatomic electron current}
\label{Sec_InterCurr}

During the time evolution of the electronic wave packet, the electron charge is redistributed both between atoms in the system and within each individual atom, thus, giving rise to the interatomic and intraatomic contributions to the probability current density $\mathbf j(\mathbf r, t_p)$, which is given by the relation 
\begin{align}
\mathbf j(\mathbf r, t_p) =\frac{i}{2}\text{Tr}\lf\{\hat \rho^{\text{m}}(t_p)\lf([\boldsymbol\nabla\hat\psi^\dagger]\hat\psi-\hat\psi^\dagger[\boldsymbol\nabla\hat\psi]\rt) \rt\} \label{ProbCurrDen}.
\end{align}
The field annihilation (creation) operator can be expanded in terms of one-particle wave functions as $\hat\psi(\mathbf r) = \sum_{\alpha}\hat c_{\alpha}\phi_\alpha(\mathbf r)$ [$\hat\psi^\dagger(\mathbf r) = \sum_{\alpha}\hat c^\dagger_{\alpha}\phi^*_\alpha(\mathbf r)$], where $\hat c_{\alpha}(\hat c_{\alpha}^{\dagger})$ annihilates (creates) a particle in one-particle state $\phi_\alpha(\mathbf r)$. Representing the one-particle wave functions as linear combinations of functions $\widetilde \phi_i(\mathbf r - \mathbf R_C)$ centered at site $ \mathbf R_C$, 
\begin{align}
\phi_{\alpha}(\mathbf r) = \sum_C\sum_i\gamma_{\alpha Ci}\widetilde\phi_i(\mathbf r - \mathbf R_C),\label{onepart}
\end{align}
the interatomic current between atoms $C_q$ and $C_r$ is
\begin{align}
&\mathbf j_{qr}(t_p) = \Im\Biggl (\sum_{I,K}\mathcal I_{IK}(t_p)\sum_{\alpha,\beta}\la\Phi_K|\hat c_{\beta}^\dagger\hat c_{\alpha}|\Phi_I\ra\label{intercurr}\\
&\quad\times\sum_{i,k}\gamma_{\beta C_r k}^*\gamma_{\alpha C_q i } \int d^3 r\widetilde\phi_k^* (\mathbf r - \mathbf R_{C_r})\boldsymbol\nabla \widetilde \phi_i(\mathbf r - \mathbf R_{C_q}) \Biggr)\nonumber.
\end{align}
Note that the current $\mathbf j_{rq}$ is opposite to $\mathbf j_{qr}$, which is consistent with the fact that the current from atom $C_q$ to $C_r$ is opposite to the current from atom $C_r$ to $C_q$. 

The interatomic electron current can be accessed via the imaginary part of the Fourier transform of a scattering pattern from $\mathbf Q$ space to real space,
\begin{align}
&\Im\lf(\frac{1}{(2\pi)^3}\int d^3 Q  \frac{dP(\mathbf Q)}{d\Omega}e^{-i\mathbf Q\cdot \mathbf r}\rt)\label{ImFurTr} \\
&\quad\propto \sum_{C_q,C_r}\mathcal J_{qr} (t_p)\delta[\mathbf r -( \mathbf R_{C_q}-\mathbf R_{C_r})]\nonumber,
\end{align}
which is the sum of delta peaks weighted by functions $\mathcal J_{qr} (t_p) = \Im(A_{qr}(t_p))$ and centered at positions corresponding to vectors connecting scattering atoms $C_q$ and $C_r$ \cite{PopovaGorelova15}. Applying Eq.~(\ref{onepart}) to dipole matrix elements $\mathbf D_{J_{C_q}I}$ and $\mathbf D_{KJ_{C_r}}$, the functions $\mathcal J_{qr} (t_p)$ can be represented as
\begin{align}
&\mathcal J_{qr} (t_p)=H_{qr}\Im\lf (\sum_{I,K}\mathcal I_{IK}(t_p)\sum_{\alpha,\beta}\la\Phi_K|\hat c_{\beta'}^\dagger \hat c_{\beta} |\Phi_{J_{C_r}}\ra \rt.\label{Jqr_via_onepart}\\
&\quad\qquad\times\lf.  \la \Phi_{J_{C_q}}|\hat c_{\alpha}^\dagger     \hat c_{\alpha'}|\Phi_I\ra\rt)\sum_{i,k}\gamma^*_{\beta C_r k} \gamma_{\alpha C_qi}\, d_{k,C_r}^*d_{i,C_q}\nonumber,
\end{align}
where 
\begin{align}
d_{i(k),C_{q(r)}} = \int d^3r \widetilde\phi_{i(k)}(\mathbf r-\mathbf R_{C_{q(r)}})&(\boldsymbol\epsilon_\i\cdot\mathbf r)\label{dipintergrals}\\
&\times\widetilde\phi^*_{\text{core}}(\mathbf r-\mathbf R_{C_{q(r)}}).\nonumber
\end{align}
Here, the operator $\hat c_{\alpha'}$ creates an electron {\it hole} in a core shell of atom $C_{q}$. Since the orbitals of electrons in core shells of heavy atoms are strongly localized, a single function $\widetilde\phi_{\text{core}}^*(\mathbf r - \mathbf R_{C_{q}})$ with the corresponding coefficient $\gamma^*_{\alpha'C_qi'}=1$ in the sum in Eq.~(\ref{onepart}) describes the core-hole wave function. The operator $\hat c_{\alpha}^\dagger$ annihilates an electron hole with a wave function $\sum_{i,C}\gamma_{\alpha C i}^*\widetilde\phi_{i}^*(\mathbf r - \mathbf R_{C})$ in the valence band, where the dynamics is taking place. We took into account that an integral $\int d^3r\widetilde\phi_{i}(\mathbf r - \mathbf R_{C}) (\boldsymbol\epsilon_\i\cdot\mathbf r)\widetilde\phi^*_{\text{core}}(\mathbf r-\mathbf R_{C_q})$ can be very well approximated by $d_{i,C_q}\delta_{\mathbf R_{C_q},\mathbf R_C} $, since the function $\widetilde\phi^*_{\text{core}}(\mathbf r-\mathbf R_{C_q})$ is strongly localized (here, $\delta_{\mathbf R_C,\mathbf R_{C_q}}$ is the Kronecker delta). Therefore, the wave function of the valence electron hole contributes only functions $\sum_{i}\gamma_{\alpha C_{q}i}^*\widetilde\phi_{i}^*(\mathbf r - \mathbf R_{C_{q}})$ to the integrals entering the dipole matrix element $\mathbf D_{J_{C_q}I}$ (analogously for the operator $\hat c_{\beta'} ^\dagger\hat c_{\beta}$ and atom $C_r$).

The probe pulse is resonant with the transition of an electron from a core shell to the valence band, where the dynamics is taking place. Therefore, the terms in the sum over $\alpha,\beta$ in Eq.~(\ref{Jqr_via_onepart}) are nonzero only if electron holes in the valence bands exist. Also, there would be no electron dynamics and, consequently, no current in the valence bands, if they are filled. Therefore, the terms in  Eqs.~(\ref{intercurr}) and (\ref{Jqr_via_onepart}) are nonzero for the same $\alpha$ and $\beta$, and factors $\gamma^*_{\beta,C_r k} \gamma_{\alpha,C_qi} $ entering factors $\mathcal J_{qr}$ and interatomic electron {\em hole} current $\mathbf j_{qr}$ are the same under the assumption stated. That means that $\mathcal J_{qr}(t_p)$ is proportional at any $t_p$ to some projection of $\mathbf j_{qr}(t_p)$ on the direction of the unit vector $\mathbf n$, as long as the ratio between $d_{i,C_q}d_{k,C_r}^*$ and $ \int d^3 r\widetilde\phi_k^* (\mathbf r - \mathbf R_{C_r})(\boldsymbol\nabla\cdot\mathbf n) \widetilde \phi_i(\mathbf r - \mathbf R_{C_q})$ is equal for every $i$ and $k$.

\section{Coherent electron-hole dynamics in potassium bromide}

\label{Section_KBr}

\begin{figure}[t]
\includegraphics[width=0.45\textwidth]{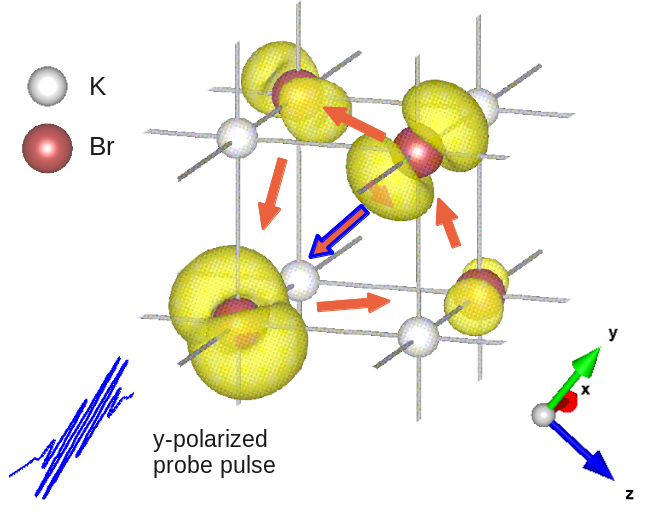}
 \caption{Electron hole density at time $t_p$ in a fragment of (KBr)$_{108}$ cluster visualized using VESTA \cite{MommaJAC11}. The orange arrows represent the electron hole currents $\mathbf j_{qr}(t_p)$ between the Br atoms, and their lengths are proportional to $|\mathbf j_{qr}(t_p)|$. The blue-framed arrow represents a current parallel to the pulse polarization.}
\label{Fig_DenKBr}
\end{figure}

As the first example we consider electron hole motion in KBr crystal. KBr is an ionic crystal with the rock-salt crystal structure, where the $4s$ electrons of K atoms are transferred to Br atoms \cite{WertheimPRB95}. The $p$-character electrons centered on the Br atoms form the outermost valence band of KBr. We assume that a pump pulse ionizes KBr by removing electrons from this band, {\it i.~e.}, by creating $p$-type electron holes centered on Br atoms, and coherently triggered their dynamics (see Fig.~\ref{Fig_DenKBr}). Generally, each electron hole would be delocalized and distributed over many Br atoms in some region. We assume that the concentration of the electron holes is sufficiently low to consider these regions isolated and the holes noninteracting. 

The second pulse, which comes after the pump at some time $t_p$, probes the electron dynamics via the ultrafast RXS, which is a two-step process. First, it resonantly excites the transition of an electron from the $1s$ shell of Br to the valence band, where the dynamics is taking place. Thereby, this absorption step directly depends on the electronic wave packet state at the time of measurement. Then, the created electron hole in the $1s$ shell of Br would be filled by either a valence electron or by an inner-shell electron of Br accompanied by the emission of a photon that reaches the detector. The second process does not directly depend on the wave packet state at time $t_p$, since the wave packet has already been destroyed in the absorption step. However, since the scattering probability is determined by both processes, absorption and emission, the scattering pattern still contains information about the wave packet.

\subsection{Computational details}

We simulated the region of the ionized KBr crystal, where a single electron hole is present, by a cubic KBr cluster consisting of 216 atoms. We performed the calculation of the electronic structure of the neutral cluster within the Hartree-Fock approach using the ab initio quantum chemistry software package MOLCAS \cite{KarlstromPSSD03} with the STO-3G basis set \cite{PietroInCh80, EMSLDataBase}, which already gives satisfactory results for KBr. The calculation was performed without periodic boundary conditions and in the real space, which allowed extracting the spatial orbitals of the singly-ionized KBr cluster and their binding energies within Koopmans' theorem \cite{KoopmansPhysica34}. There are 324 valence orbitals in the cluster, since there are six $p$-like valence electrons per Br atom. Therefore, there are 324 electronic states involved in the dynamics of the wave packet in our simulation.


We assume that the pump and probe pulses do not overlap temporally. Therefore, their actions can be described separately \cite{SantraPRA11}.  We further assume that the pump pulse has created a perfectly coherent superposition of the electronic states, so that the elements $\mathcal I_{IK}(t)$ in Eq.~(\ref{El_den}) can be represented as $\mathcal I_{IK}(t)=C_IC_K^*e^{-i(E_I-E_K)t}$, where $C_I$ and $C_K$ are time-independent coefficients. In principle, these coefficients must be determined from the description of the specific process triggering the wave packet. But in this paper, our goal is to demonstrate how, using ultrafast RXS, one can extract information about nonstationary electron dynamics in a crystal independently from how this electron dynamics was excited. Therefore, we do not concentrate on the pump process and choose coefficients $C_{I,K}$ randomly, since the conclusions presented below do not depend on the specific set of the elements $\mathcal I_{IK}(t)$ and are expected to remain valid even if the state of the electronic system is a statistical mixture of states (i.~e., $\mathcal I_{IK}(t)\ne C_IC_K^*e^{-i(E_I-E_K)t}$). 

The distribution of 324 energy levels reflects the main features of the density of states of the valence band of bulk KBr crystal \cite{WertheimPRB95}. With increasing number of atoms in the cluster, the intervals between the energy levels decrease and the distribution of the states tend to the density of states of the valence band. Since the behavior of a wave packet is determined by the energy states involved in the dynamics, the description of the wave packet dynamics in the crystal would improve with increasing number of atoms in the cluster. Our focus here is on analyzing time-resolved scattering patterns from a wave packet in a periodic structure. As will be discussed below, the conclusions about properties of such scattering patterns do not change with increasing cluster size.


\subsection{Scattering pattern}
\label{Subsec_ScatterinPatternKBr}

\begin{figure}[t]
\includegraphics{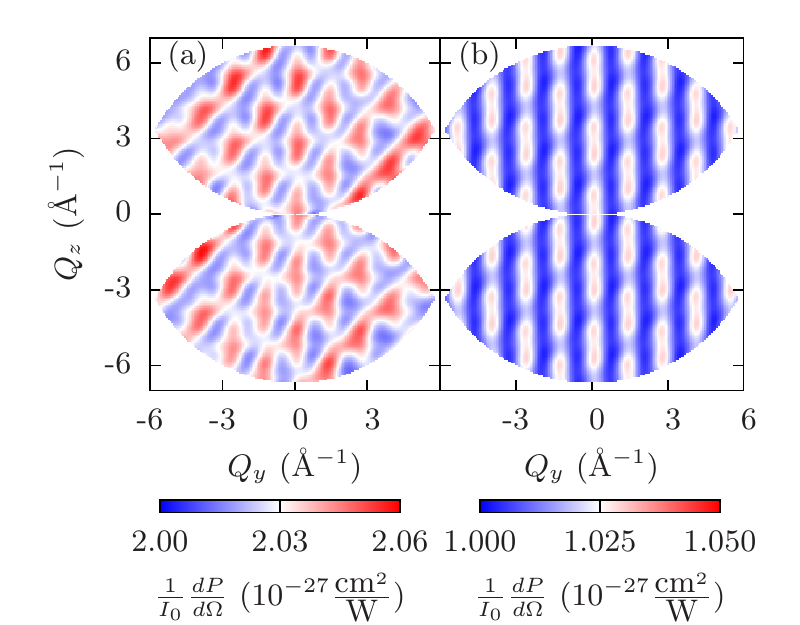}
  \caption{Scattering patterns at the probe-pulse intensity $I_0$ from the ionized KBr cluster obtained by a $y$-polarized x-ray pulse. (a) No polarization filter is applied in the measurement of a scattered photon. (b) A polarization filter transmitting $y$-polarized scattered photons is applied and the pattern is divided by the function $g(\mathbf Q)$ [Eq.~(\ref{PolarFilter})]. }
\label{Fig_KBrpatt}
\end{figure}

The duration of the probe pulse of 200 as is chosen such that it is sufficiently short to capture the electron dynamics of the wave packet, the characteristic time scale of which is determined by the bandwidth of the valence band of 2.5 eV \cite{WertheimPRB95}. The photon energy of the probe pulse $\omega_\i\approx13$ keV is tuned to the $K$ edge of Br, providing a spatial resolution of 0.9 \r A. Such sub-femtosecond hard x-ray pulses can be produced at free-electron lasers using the strategies described in Refs.~\onlinecite{ZholentsPRL04, EmmaPRL04, SaldinPhysRevSTAB06, KumarAppSciences13, TanakaPRL13,PratPRL15}. The probe pulse propagates along the $x$ direction and its polarization is along the $y$ axis, parallel to one of the vectors connecting two Br atoms (see Fig.~\ref{Fig_DenKBr}). 

It follows from Eq.~(\ref{DSP_DipApp_extended}) that photons emitted from localized electrons filling the $1s$ core hole in Br do not provide a structure-dependent contribution to the scattering pattern. Namely, if an electron localized at some atom $C_l$ would take part in the emission step by filling the core hole, then the final state $|F\ra$ would be a state with an electron hole localized at $C_l$. As a result, all terms $(\mathbf D_{J_{C_r}F}\cdot \boldsymbol\epsilon_\s)(\mathbf D_{FJ_{C_q}}\cdot \boldsymbol\epsilon_\s^*)$ except for the term $(\mathbf D_{J_{C_l}F}\cdot \boldsymbol\epsilon_\s)(\mathbf D_{FJ_{C_l}}\cdot \boldsymbol\epsilon_\s^*)$ in the sum over scattering atoms $C_q$ and $C_r$ in Eq.~(\ref{DSP_DipApp_extended}) would be zero. Thus, photons emitted by electrons from localized orbitals filling the core hole would provide a signal that is constant over $\mathbf Q$ space, thereby, contributing only to the background. These photons have energies lower than energies of transitions of electrons from the outermost valence band to the $1s$ shell of Br and can be suppressed by the spectral window function $W(\omega_{\k_\s})$ centered at $\omega_\i$. Thus, we assume that the spectral window function $W(\omega_{\k_\s})$ is centered at $\omega_\i$ and suppresses all photons emitted by electrons lying deeper than the outermost valence band. Without this assumption, the signal in a scattering pattern would be larger, but its contrast would be lower.

Figure~\ref{Fig_KBrpatt}a shows the scattering pattern in the $Q_yQ_z$ plane at $Q_x=0$ from the singly ionized KBr cluster obtained by the probe pulse arriving at time $t_p$. The sample must be rotated about the $y$ axis as described in Appendix \ref{App_Rotkin} in order to acquire data in this plane. Two conditions must be satisfied for $Q$ points in the plane to be accessible. The first condition is that $|\mathbf Q|$ cannot exceed $2|\k_\i|\sin(\theta_{\text{max}}/2)$, where $\theta_{\text{max}}$ is the maximum scattering angle, which we set to $60^{\circ}$. The second condition is due to the inexistence of a configuration of vectors $\mathbf k_\i$ and $\mathbf k_\s$ for some $Q$ points (see Appendix \ref{App_Rotkin} and Refs.~\onlinecite{HoJCP09, SaldinActaCryst10} for details). This condition limits the accessible area by two circles of radius $|\k_\i|$ centered at the points $(Q_y,Q_z)=(0,\pm|\k_\i|)$.

The $\mathbf Q$ dependence of the scattering pattern in Fig.~\ref{Fig_KBrpatt}a is given not only by the trigonometric functions in Eq.~(\ref{DSP_DipApp_extended}), but also by the function $H_{qr}$, which contains a sum over the two independent polarizations of the scattered photon. Since the vector $\mathbf k_\s$ is different for every $Q$ point, the terms $\sum_{F,s_\s}(\mathbf D_{FJ_{C_q}}\cdot \boldsymbol\epsilon_{\s}^*)(\mathbf D_{FJ_{C_r}}\cdot \boldsymbol\epsilon_{\s})$ in the function $H_{qr}$ are also different and depend on $\mathbf Q$. This additional $\mathbf Q$ dependence will manifest itself in the Fourier transform of the scattering pattern to the real space, which would complicate its analysis. In order to eliminate this dependence, we apply a polarization filter that transmits scattered photons with polarization $\boldsymbol \epsilon_p$ (see Appendix \ref{App_polarfilter}). Then, the term $\sum_{F,s_\s}(\mathbf D_{FJ_{C_q}}\cdot \boldsymbol\epsilon_{\s}^*)(\mathbf D_{FJ_{C_r}}\cdot \boldsymbol\epsilon_{\s})$ will be substituted by $g(\mathbf Q) \sum_F(\mathbf D_{FJ_{C_q}}\cdot\boldsymbol \epsilon_p)(\mathbf D_{J_{C_r}F}\cdot\boldsymbol \epsilon_p^*)$, where function the $g(\mathbf Q)$ is given in Eq.~(\ref{PolarFilter}). Thus, by applying a polarization filter, the dependence of the factor $H_{qr}$ on $\mathbf Q$ can be factored out independently of which polarization $\boldsymbol\epsilon_p$ the filter transmits. 

Figure~\ref{Fig_KBrpatt}b shows the scattering pattern at $t=t_p$ obtained with a filter transmitting only $y$-polarized photons and divided by the function $g(\mathbf Q)$. Now, the pattern contains only the $\mathbf Q$ dependence determined by the trigonometric functions in Eq.~(\ref{DSP_DipApp_extended}). In this way the periodicity of the pattern can be straightforwardly determined, and it can be extrapolated to a region with inaccessible $\mathbf Q$ in order to perform the Fourier transform of the pattern to the real space. Looking at the pattern carefully, one can notice that the point at $(Q_y,Q_z) = (0,0)$ is not a center of inversion symmetry, corroborating the discussion above that the pattern is not determined by the usual structure factor squared. The shape and positions of the peaks in the pattern change depending on the probe-pulse arrival time. The contrast and the mean signal of the pattern also slightly change in time.

Note that this scattering pattern is formed by a single hole distributed over Br atoms in the cluster, and the diffraction peaks in the pattern are due to the interference of segments of the electron hole centered on different atoms (see Fig.~\ref {Fig_DenKBr}). With increasing cluster size, the contrast and the mean signal strength of a scattering pattern averaged over time will not change, since still only a single electron hole would contribute to the signal. A real scattering pattern from an ionized KBr crystal would be formed by the sum of signals from all electron holes in the crystal, which would enhance the number of photons that reach the detector by a corresponding factor. In Appendix \ref{App_RequiredPhotons}, we estimated that a KBr crystal with one electron hole per one thousand atoms has to be irradiated with $3\times 10^{12}$ photons in order to achieve a signal of one photon per pixel on average. Additionally, the contrast would be enhanced due to Bragg reflections from electron holes that move coherently. However, scattering patterns from KBr crystal and other alkali halides would always have a low contrast compared to other systems, since valence electrons in alkali halides are to a large degree localized \cite{WertheimPRB95}.


\subsection{Fourier transform of the scattering pattern}

\begin{figure}[t]
\includegraphics{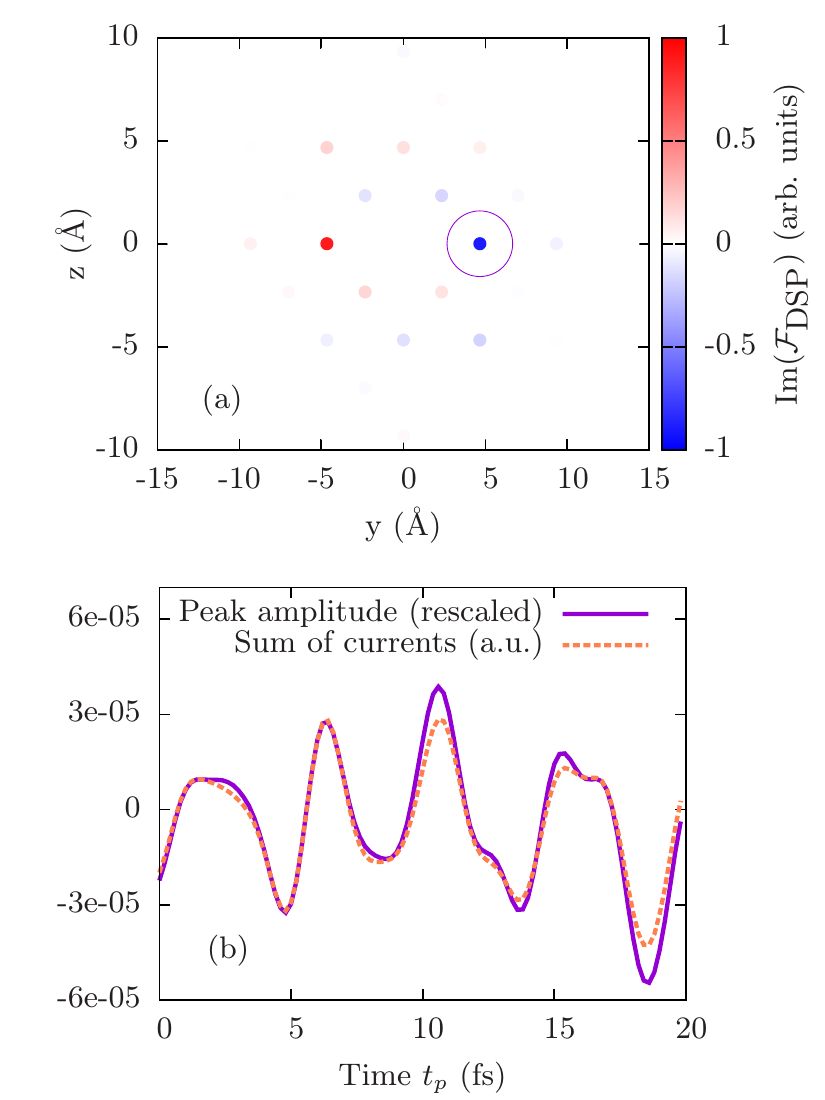}
  \caption{(a) The Fourier transform from $\mathbf Q$ space to real space of the scattering pattern in Fig.~\ref{Fig_KBrpatt}b. (b) Solid violet line: time evolution of the amplitude of the circled peak in panel (a). Orange dashed line: the sum of the currents between pairs of atoms connected by the vector $(0,R_{\text{Br-Br}},0)$. }
\label{Fig_CurrFurKBr}
\end{figure}

\begin{figure}[t]
\includegraphics{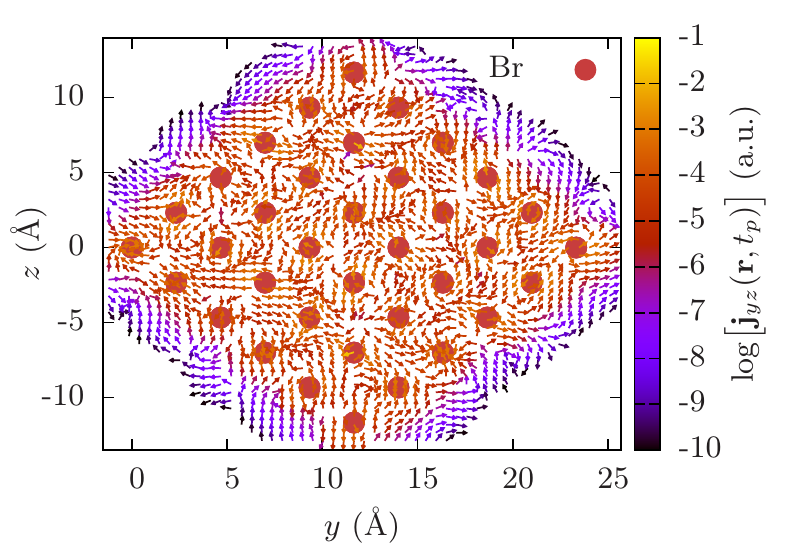}
  \caption{Br atoms in the projection of the KBr cluster on the $yz$ plane, illustrating how the KBr cluster is ``seen" by the resonant x-ray probe pulse propagating in the $x$ direction. Also shown is the projection of the probability current density [Eq.~(\ref{ProbCurrDen})] on the $yz$ plane, $\mathbf j_{yz}(\mathbf r, t_p)$ .}
\label{Fig_gradKBr}
\end{figure}

Figure~\ref{Fig_CurrFurKBr}a shows the imaginary part of the Fourier transform from $\mathbf Q$ space to real space of the pattern from Fig.~\ref{Fig_KBrpatt}b. This and further plots of the imaginary part of the Fourier transform we obtain with Eq.~(\ref{ImFurTr}), which provides the ``ideal" Fourier transform of scattering patterns, as if they were extrapolated to the region of infinite $Q_y$ and $Q_z$. The Fourier transform consists of delta peaks at points corresponding to vectors connecting pairs of Br atoms (see Fig.~\ref{Fig_gradKBr}). For instance, the peak outlined by the circle in Fig.~\ref{Fig_CurrFurKBr}a corresponds to the vector $(0,R_{\text{Br-Br}},0)$ that connects all nearest-neighbor Br atoms lying in the $y$ direction, where $R_{\text{Br-Br}} = 4.7$ \r A is the distance between the nearest-neighbor Br atoms. Amplitudes of peaks at points $(R_y,R_z)$ are opposite to amplitudes of peaks at $(-R_y,-R_z)$. 

Now let us consider the correspondence between the imaginary part of the Fourier transform and the interatomic electron current, which we discussed in Sec.~\ref{Sec_InterCurr}. The electron hole is created in the valence band of KBr that is formed by $4p$-type orbitals centered on Br atoms. Therefore, the atomic functions $\widetilde\phi_i(\mathbf r - \mathbf R_C)$, which we use to expand the wave function of the electron hole [see Eq.~(\ref{onepart})], are simply $4p_x$-, $4p_y$- and $4p_z$-type orbitals centered on site $C$, which we designate as $\widetilde\phi_{p_x}$, $\widetilde\phi_{p_y}$ and $\widetilde\phi_{p_z}$, respectively. According to Eq.~(\ref{intercurr}), the interatomic electron hole current is given by
\begin{align}
&\mathbf j_{qr}(t_p) =\Im\Biggl(\sum_{I,K}\mathcal I_{IK}(t_p) \label{InterCurrKBr}\\ 
&\quad\times\sum_{b,c}\gamma^{*K}_{C_rp_b}\gamma^I_{C_qp_c}  \int d^3r\widetilde\phi_{p_b}^*(\mathbf r-\mathbf R_r)\boldsymbol \nabla\widetilde\phi_{p_c}(\mathbf r-\mathbf R_q)\Biggr),\nonumber
\end{align} 
where $b$ and $c$ stand for $x$, $y$ and $z$ and $\gamma^I_{C_qp_c}$ and $\gamma^{K}_{C_rp_b}$ are coefficients for which the matrix element $\la\Phi_K|\hat c_{\beta}^\dagger\hat c_{\alpha}|\Phi_I\ra$ in Eq.~(\ref{intercurr}) is nonzero. Let atoms $C_q$ and $C_r$ be nearest-neighbor atoms aligned along the $y$ direction. Then, the integral $\int d^3r\widetilde\phi_{p_y}^*(\mathbf r-\mathbf R_r)\nabla_y\widetilde\phi_{p_y}(\mathbf r-\mathbf R_{q})$ is much larger than the other integrals involving functions $\widetilde\phi_{p_x}$, $\widetilde\phi_{p_y}$, $\widetilde\phi_{p_z}$ and operators $\nabla_x$, $\nabla_y$, $\nabla_z$, meaning that the $y$ component of the current $\mathbf j_{qr}$ would be the dominating one. Since only one integral provides the dominating contribution to the amplitude of $\mathbf j_{qr}$, and the $4p_y$-type function is real, $|\mathbf j_{qr}(t_p)|$ is simply given by 
\begin{align}
|\mathbf j_{qr}(t_p)| \approx &\lf(\int d^3r\widetilde\phi_{p_y}(\mathbf r-\mathbf R_r)\nabla_y\widetilde\phi_{p_y}(\mathbf r-\mathbf R_q)\rt)\\
&\qquad\qquad\times\Im\Biggl(\sum_{I,K}\mathcal I_{IK}(t_p) \gamma^{*K}_{C_rp_y}\gamma^I_{C_qp_y}  \Biggr).\nonumber
\end{align}
We obtain the same result, that interatomic electron-hole currents between nearest-neighbor Br atoms in KBr are aligned along vectors connecting the corresponding atoms (see Fig.~\ref{Fig_DenKBr}), by numerically calculating the currents. Also, we find that the current between atoms not lying next to each other, is negligible. This is because the functions  $\widetilde\phi_{p_y}(\mathbf r-\mathbf R_r)$ and $\nabla_y\widetilde\phi_{p_y}(\mathbf r-\mathbf R_q)$ centered on atoms $C_r$ and $C_q$ that are far apart, do not overlap with each other.

Now let us consider the factor $\mathcal J_{qr}$, which, according to Eq.~(\ref{Jqr_via_onepart}), is
\begin{align}
&\mathcal J_{qr}(t_p) = H_{qr}|d_{p_y}|^2 \Im\Biggl(\sum_{I,K}\mathcal I_{IK}(t_p) \gamma^{*K}_{C_rp_y}\gamma^I_{C_qp_y} \Biggr)\nonumber \\
&d_{p_y}=  \int d^3r \widetilde\phi_{p_y}(\mathbf r-\mathbf R_q) y \widetilde \phi_{1s}^*(\mathbf r-\mathbf R_q)\\
&\phantom{d_{p_y}}=\int d^3r \widetilde\phi_{p_y}(\mathbf r-\mathbf R_r) y \widetilde \phi_{1s}^*(\mathbf r-\mathbf R_r)\nonumber,
\end{align} 
where $\widetilde \phi_{1s}(\mathbf r-\mathbf R_{q(r)})$ is the wave function of an electron hole in the $1s$ shell of the Br atom at site $C_{q(r)}$. Here, we took into account that the integrals $d_{p_x}$ and $d_{p_z}$ are zero for a $y$-polarized probe pulse. Comparing functions $\mathcal J_{qr}(t_p)$ and $|\mathbf j_{qr}(t_p)|$, one can see that they are proportional at every $t_p$ with the coefficient of proportionality being given by $H_{qr}|d_{p_y}|^2/\int d^3r\widetilde\phi_{p_y}(\mathbf r-\mathbf R_r)\nabla_y\widetilde\phi_{p_y}(\mathbf r-\mathbf R_q)$.  Due to the symmetry of the KBr crystal, which has the rock-salt crystal structure, all pairs of Br atoms connected by the same vector are equivalent to each other. Thus, the sum $\sum_F(\mathbf D_{FJ_{C_{q}}}\cdot\boldsymbol\epsilon_\s^*)(\mathbf D_{J_{C_{r}F}}\cdot\boldsymbol\epsilon_\s)$ and, consequently, the factor $H_{qr}$ are equal for all pairs of Br atoms connected by the same vector $\Delta\mathbf R_{qr} = \mathbf R_r-\mathbf R_q$. The integral $\int d^3r\widetilde\phi_{p_y}(\mathbf r-\mathbf R_r)\nabla_y\widetilde\phi_{p_y}(\mathbf r-\mathbf R_q)$ also does not change for the same $\Delta\mathbf R_{qr}$. Thus, the coefficient of proportionality is identical for all pairs of Br atoms connected by equal vectors.

The factor $\mathcal J_{qr}$ corresponding to the two atoms at sites $C_q$ and $C_r$ connected by the vector $(0,R_{\text{Br-Br}},0)$ contributes to the amplitude of the peak outlined by the circle in Fig.~\ref{Fig_CurrFurKBr}a. This peak is formed not only by the single factor $\mathcal J_{qr}$, but by the sum of all factors $\mathcal J_{q'r'}$, the corresponding pair of atoms $C_{q'}$ and $C_{r'}$ being connected by vectors with the projection on the $yz$ plane equal to $(R_{\text{Br-Br}},0)$. This means that not only nearest-neighbor Br-atom pairs lying in the $y$ direction, but also Br-atom pairs separated by a vector $(2NR_{\text{Br-Br}},R_{\text{Br-Br}},0)$, where $N$ is an integer, contribute to this peak. However, factors $\mathcal J_{q'r'}$ for atoms $C_{q'}$ and $C_{r'}$ not lying next to each each other, are negligible due to the term $\sum_F(\mathbf D_{FC_{q'}}\cdot\boldsymbol\epsilon_\s^*)(\mathbf D_{FC_{r'}}\cdot\boldsymbol\epsilon_\s)$ entering $\mathcal J_{q'r'}$ [see Eq.~(\ref{DSP_DipApp_extended}) and (\ref{Jqr_via_onepart})]. The less spatial orbitals are distributed among both atoms $C_{q'}$ and $C_{r'}$, the more this term decreases. Thus, the peak at $(R_{\text{Br-Br}},0)$ in Fig.~\ref{Fig_CurrFurKBr}a is formed just by the sum of the factors $\mathcal J_{q'r'}$ corresponding to the pairs of atoms lying next to each other along the $y$ direction and is proportional to the sum of the interatomic electron hole currents between them.

In Fig.~\ref{Fig_CurrFurKBr}b, we depict the amplitude of the encircled peak as a function of the time of the probe pulse arrival and the time evolution of the computed sum of the currents between all pairs of nearest-neighbor Br atoms connected by the vector $(0,R_{\text{Br-Br}},0)$. In agreement with the previous discussion, we find that the time evolution of the amplitude of this peak precisely follows the time evolution of the sum of the currents.

To sum up, the interatomic electron currents in the outermost valence band of KBr are nonzero for nearest-neighbor Br atoms and are aligned along directions connecting the pairs of atoms. The sum of the interatomic electron currents between atoms lying along a vector $\Delta\mathbf R_{qr}$ is encoded in the ultrafast RXS scattering pattern obtained by a probe pulse polarized along $\Delta\mathbf R_{qr}$. The time evolution of the peak at the position $\Delta\mathbf R_{qr}$ in the imaginary part of the Fourier transform of the scattering pattern follows the time evolution of the sum of interatomic currents. This concept works precisely for systems where the same single function $\widetilde \phi_i$ per scattering atom contributes to both the interatomic electron currents in some direction $\Delta\mathbf R_{qr}$ and the signal due to the probe pulse polarized along some direction $\boldsymbol\epsilon_\i$ (in general, not necessarily the same as $\Delta\mathbf R_{qr}$). This conclusion is deduced for electronic wave packets in periodic structures, but also works for the finite KBr cluster as follows from Fig.~\ref{Fig_CurrFurKBr}b.

\section{Coherent electron-hole dynamics in Germanium}
 
\label{Section_Ge} 
 
\begin{figure}[b]
\includegraphics[width=0.40\textwidth]{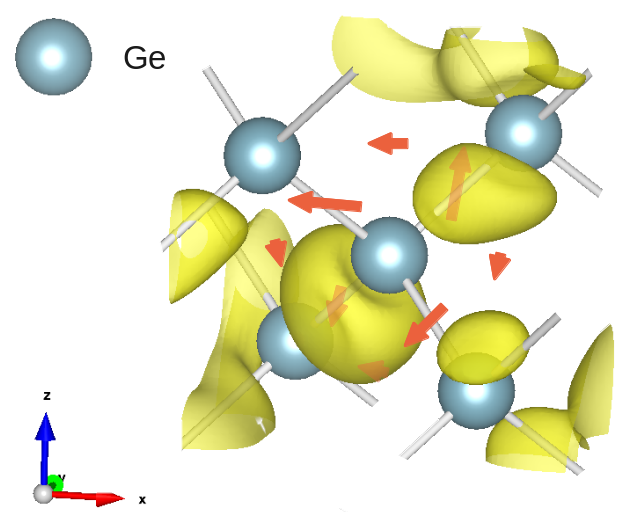}
 \caption{Electron hole density at time $t_p$ in a fragment of Ge cluster visualized using VESTA software \cite{MommaJAC11}. The orange arrows represent the electron hole currents $\mathbf j_{qr}(t_p)$ between the Ge atoms, and their lengths are proportional to $|\mathbf j_{qr}(t_p)|$.}
\label{Fig_DenGe}
\end{figure}

The second example that we consider is the coherent valence electron hole motion in germanium crystal launched by a photoionizing pump pulse (see Fig.~\ref{Fig_DenGe}). We assume a probe pulse with a photon energy $\omega_\i\approx 11$ keV, which corresponds to a spatial resolution of 1.1 \r A, tuned to the transition of an electron from the $1s$ shell of Ge to the valence band. The probe pulse duration of 200 as is the same as in the previous example. We again assume that the electron holes in the crystal do not interact and simulate an isolated region of Ge crystal, where a single electron hole is present, by a cluster of 83 Ge atoms. The shape of the cluster is chosen such that there is a maximum number of nearest-neighbor atoms per atom. Therefore, the boundary of this cluster tends to a sphere. Our results do not depend on the choice of cluster shape, but do depend on the number of nearest-neighbor atoms, since they provide the largest contributions to a scattering pattern.

We performed the calculation of the electronic structure of the neutral Ge cluster within the Hartree-Fock approach using the ab initio quantum chemistry software package MOLCAS \cite{KarlstromPSSD03} with the correlation-consistent basis set, cc-pVDZ \cite{EMSLDataBase} without periodic boundary conditions. We obtained the spatial orbitals of the singly-ionized Ge cluster and their binding energies within Koopmans' theorem \cite{KoopmansPhysica34}. Germanium has the diamond crystal structure, and, similarly to diamond, each Ge atom forms covalent $sp^3$ bonds to four neighboring Ge atoms. Since Ge atoms on the surface of the cluster have less than four nearest neighbors, they have unsaturated bonds, which lead to a distortion of the electronic structure of the whole cluster. We solved this problem by saturating these bonds with hydrogen atoms. This does not influence the results, since only Ge atoms scatter in our case.

The $sp^3$-hybridized electrons form two outermost valence bands of Germanium with a maximum energy splitting of 4.5 eV \cite{EastmanPRB74}. The wave functions $\widetilde\phi_{1(2),i}$ of the hybrid orbitals are a linear combination of one $s$ and three $p$ orbitals of each Ge atom, which we denote as $\widetilde \phi_s$, $\widetilde \phi_{p_x}$, $\widetilde \phi_{p_y}$ and $\widetilde \phi_{p_z}$, respectively. The indices 1 and 2 in the function $\widetilde\phi_{1(2)i}$ stand for each of the two atoms in the primitive unit cell of Ge and $i$ designates the index of the wave function:
\begin{align}
&\widetilde\phi_{1(2)1} = \frac12\lf [ \widetilde \phi_s \pm (\widetilde \phi_{p_x} +\widetilde \phi_{p_y} +\widetilde \phi_{p_z} ) \rt],\nonumber\\
&\widetilde\phi_{1(2)2} = \frac12\lf [ \widetilde \phi_s \pm (\widetilde \phi_{p_x} -\widetilde \phi_{p_y} -\widetilde \phi_{p_z} ) \rt],\label{sp3functions}\\
&\widetilde\phi_{1(2)3} = \frac12\lf [ \widetilde \phi_s \pm (-\widetilde \phi_{p_x} +\widetilde \phi_{p_y} -\widetilde \phi_{p_z} ) \rt],\nonumber\\
&\widetilde\phi_{1(2)4} = \frac12\lf [ \widetilde \phi_s \pm (-\widetilde \phi_{p_x} -\widetilde \phi_{p_y} +\widetilde \phi_{p_z} ) \rt],\nonumber
\end{align}
where the $\widetilde\phi_{1i}$ functions have the plus sign in front of the round brackets, and the $\widetilde\phi_{2i}$ have the minus sign. The $sp^3$ orbitals are extended in the directions of the nearest neighbors. For instance, let the atom situated at position $(0,0,0)$ have index 1. Then, $\widetilde\phi_{1i}$ functions are centered on this atom and its nearest neighbors are situated at positions $(a_0/4,a_0/4,a_0/4)$, $(a_0/4,-a_0/4,-a_0/4)$, $(-a_0/4,a_0/4,-a_0/4)$ and $(-a_0/4,-a_0/4,a_0/4)$, where $a_0=5.658$ \r A is the lattice constant of Ge. All these atoms have index 2, since functions $\widetilde\phi_{2i}$ centered on them are extended in the opposite directions as compared to $\widetilde\phi_{1i}$. Therefore, nearest-neighbor atomic orbitals in Ge always have different indices.

\begin{figure}[t]
\includegraphics{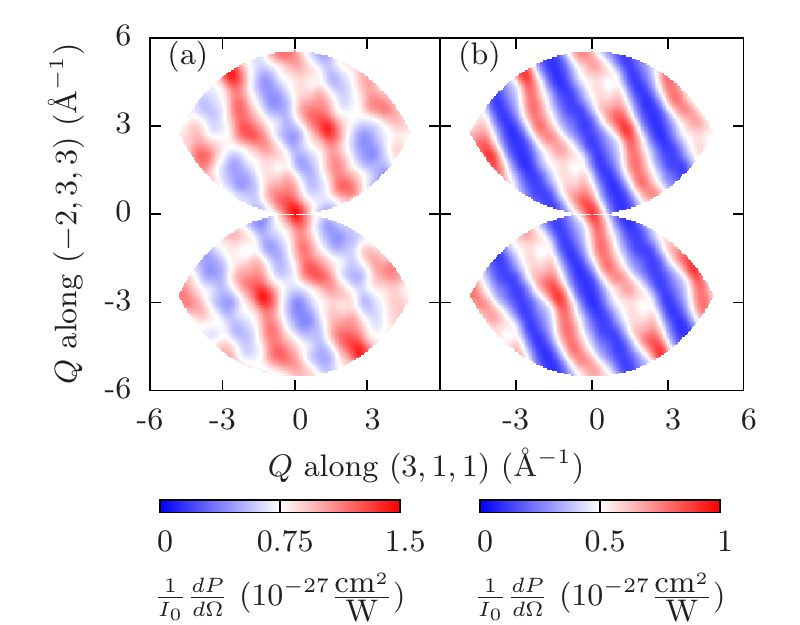}
  \caption{Scattering patterns at the probe-pulse intensity $I_0$ from the ionized Ge cluster obtained by an x-ray pulse polarized along $(3,1,1)$ direction. $Q$ along direction perpendicular to the illustrated plane is zero. (a) No polarization filter is applied in the measurement of the scattered photon. (b) Polarization filter transmitting scattered photons polarized along $(1,1,1)$ direction is applied and the pattern is divided by the function $g(\mathbf Q)$ [Eq.~(\ref{PolarFilter})]. 
  }
\label{Fig_Gepatt111}
\end{figure}

Now let us consider the electron hole currents between nearest-neighbor Ge atoms. Let the atomic orbitals at site $\mathbf R_{q}$ have index 1, and the atomic orbitals at site $\mathbf R_{r}$ have index 2, then
\begin{align}
&\mathbf j_{qr}(t_p) =\Im\Biggl(\sum_{I,K}\mathcal I_{IK}(t_p)\label{InterCurrGe}\\
&\qquad \times\sum_{i,k}\gamma^{*K}_{C_rk}\gamma^I_{C_qi}  \int d^3r\widetilde\phi_{2k}^*(\mathbf r-\mathbf R_r)\boldsymbol \nabla\widetilde\phi_{1i}(\mathbf r-\mathbf R_q)\Biggr),\nonumber
\end{align} 
where $\gamma^I_{C_qi}$ and $\gamma^{K}_{C_rk}$ are the coefficients for which the matrix element $\la\Phi_K|\hat c_{\beta}^\dagger\hat c_{\alpha}|\Phi_I\ra$ in Eq.~(\ref{intercurr}) is nonzero. This time, all integrals involving functions $\widetilde\phi_{1(2)i(k)}$ and operators $\nabla_{x,y,z}$ are of the same order, and there is no preferred direction of the interatomic currents between nearest-neighbor atoms in Ge (see Fig.~\ref{Fig_DenGe}).  

The corresponding factor $\mathcal J_{qr}(t_p)$ for a probe pulse polarized in some direction $\boldsymbol\epsilon_\i$ is
\begin{align}
&\mathcal J_{qr}(t_p) = H_{qr} \Im\Biggl(\sum_{I,K}\mathcal I_{IK}(t_p) \sum_{i,k}\gamma^{*K}_{C_rk}\gamma^I_{C_qi} d_{2k}^*d_{1i}\Biggr), \\
&d_{1(2)i(k)}=  \int d^3r \widetilde\phi_{1(2)i(k)}(\mathbf r) (\boldsymbol\epsilon_\i\cdot\mathbf r) \widetilde \phi_{1s}^*(\mathbf r) \nonumber,
\end{align}
where $\widetilde \phi_{1s}(\mathbf r-\mathbf R)$ designates the wave function of an electron hole in the $1s$ shell of the Ge atom located at position $\mathbf R$. For the factor $\mathcal J_{qr}(t_p)$ to follow the time evolution of some projection of $\mathbf j_{qr}(t_p)$ onto a unit vector $\mathbf n$, the polarization of the probe pulse should be chosen such that $d_{2k}^*d_{1i}$ has the same coefficient of proportionality with respect to $\int d^3r\widetilde\phi_{2k}^*(\mathbf r-\mathbf R_r)(\boldsymbol \nabla\cdot \mathbf n)\widetilde\phi_{1i}(\mathbf r-\mathbf R_q)$ for every $i$ and $k$.  

We find that, for interatomic currents between atoms connected by the vector $(a_0/4,a_0/4,a_0/4)$, a pair of $\mathbf n$ and $\boldsymbol\epsilon_\i$, for which the condition above is approximately satisfied, is $\mathbf n = (-\frac{2}{\sqrt6},\frac1{\sqrt6},\frac1{\sqrt6})$ and $\boldsymbol \epsilon_\i = (\frac 3{\sqrt{11}},\frac1{\sqrt {11}},\frac1{\sqrt {11}})$. In Fig.~\ref{Fig_Gepatt111}, we show the scattering patterns obtained with the probe pulse with polarization $(\frac 3{\sqrt{11}},\frac1{\sqrt {11}},\frac1{\sqrt {11}})$. We again apply a spectral window function $W(\omega_{\k_\s})$ that suppresses photons emitted by electrons lying in bands deeper that the two outermost ones. This condition is not necessary, but desirable to obtain a higher contrast in the scattering patterns. No polarization filter for scattered photons is applied in the case of Fig.~\ref{Fig_Gepatt111}a. The scattering pattern in Fig.~\ref{Fig_Gepatt111}b is obtained with a filter transmitting scattered photons polarized along $(1,1,1)$ and is divided by the function $g(\mathbf Q)$ [see Eq.~(\ref{PolarFilter})]. Here, the contrast of the scattering patterns is much higher than in the previous example (see Fiq.~\ref{Fig_KBrpatt}), since nearest-neighbor Ge atoms in the crystal are closer to each other, at a distance of 2.45 \r A, which is about two times less than the distance between nearest-neighbor Br atoms in KBr. As a consequence, the spatial orbitals in Ge are rather delocalized, leading to an increase of the factor $\sum_F(\mathbf D_{FJ_{C_q}}\cdot \boldsymbol\epsilon_{\s}^*)(\mathbf D_{J_{C_r}F}\cdot \boldsymbol\epsilon_{\s})$, as we have discussed earlier. We also find that three times less photons, $1\times10^{12}$, than in the case of KBr are necessary to obtain a signal of one photon per pixel on average in Ge crystal with one electron hole per one thousand atoms (see Appendix \ref{App_RequiredPhotons}).

\begin{figure}[t]
\includegraphics{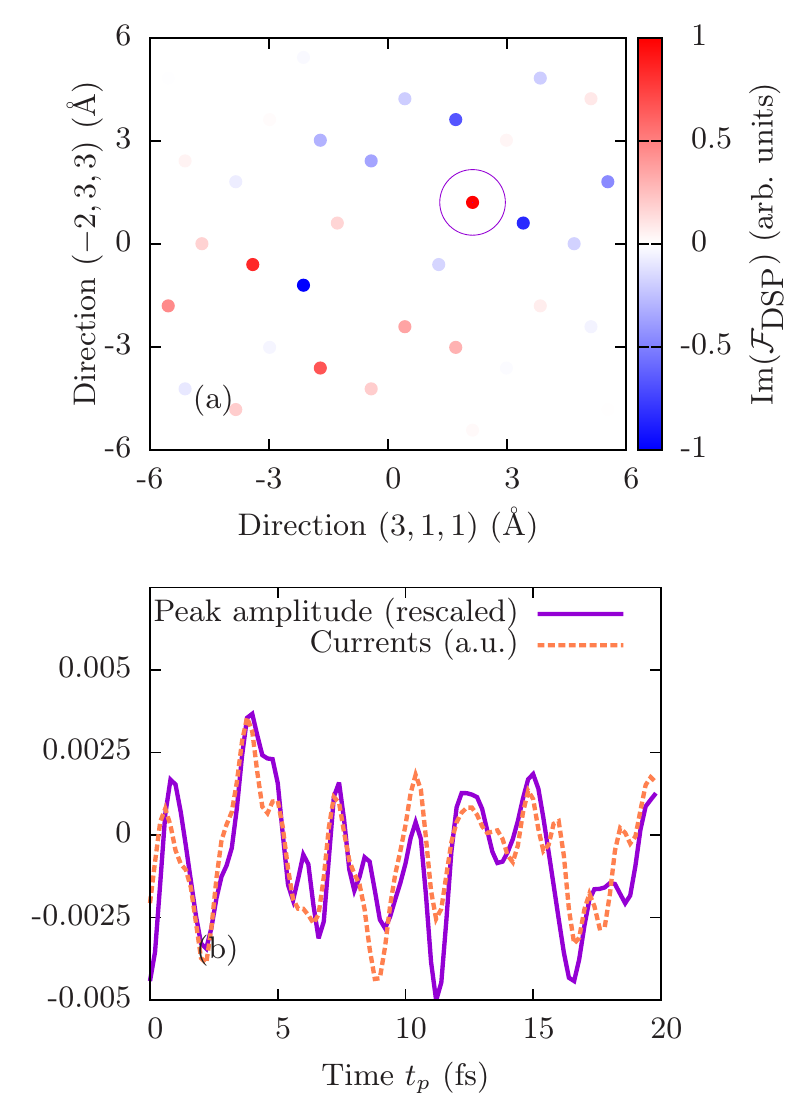}
  \caption{(a) The Fourier transform from $\mathbf Q$ space to real space of the scattering pattern in Fig.~\ref{Fig_Gepatt111}b. (b) Solid violet line: time evolution of the amplitude of the circled peak in panel (a). Orange dashed line: $\sum_{qr}(\mathbf j_{qr}\cdot\mathbf n)+\sum_{q'r'}(\mathbf j_{q'r'}\cdot\mathbf n)$, where $\Delta\mathbf R_{qr}=-\Delta\mathbf R_{q'r'} = (a_0/4,a_0/4,a_0/4)$. }
\label{Fig_CurrFurGe111}
\end{figure}

Figure \ref{Fig_CurrFurGe111}a shows the imaginary part of the Fourier transform from $\mathbf Q$ space to real space of the scattering pattern in Fig.~\ref{Fig_Gepatt111}b. This Fourier transform has much more pronounced peaks as compared to the case of KBr (see Fig.~\ref{Fig_CurrFurKBr}a), since not only nearest-neighbor atoms, but also next-nearest-neighbor atoms in Ge (distances 4 \r A) provide noticeable factors $\mathcal J_{qr}$. The peak corresponding to atoms connected by the vector $(a_0/4,a_0/4,a_0/4)$ is indicated by the circle. The plane of the scattering pattern was chosen such that this peak does not overlap with any other prominent peaks. Thus, its amplitude is given by the sum of factors $\mathcal J_{qr}$ corresponding to atoms $C_q$ and $C_r$ connected by the vector $(a_0/4,a_0/4,a_0/4)$. 

Although each individual $\mathcal J_{qr}(t_p)$ is approximately proportional to the projection of the corresponding interatomic current, $\mathbf j_{qr}(t_p)\cdot\mathbf n$, the coefficient of proportionality is different for different pairs. This is connected with the fact that a Ge crystal has two scattering atoms in the primitive unit cell in contrast to the KBr crystal, which has only one Br atom in the primitive unit cell. Namely, let atoms $C_q$ and $C_r$ have indices 1 and 2 and atoms $C_{q'}$ and $C_{r'}$ have indices 2 and 1, respectively. If both pairs of these atoms are nearest-neighbor atoms along the (1,1,1) direction, then $\Delta\mathbf R_{qr} = -\Delta \mathbf R_{q'r'}$. The integrals $\int d^3r\widetilde\phi_{2k}^*(\mathbf r)(\boldsymbol \nabla\cdot \mathbf n)\widetilde\phi_{1i}(\mathbf r-\Delta\mathbf R_{qr})$ and $\int d^3r\widetilde\phi_{1k}^*(\mathbf r-\Delta\mathbf R_{qr})(\boldsymbol \nabla\cdot \mathbf n)\widetilde\phi_{2i}(\mathbf r)$ in Eq.~(\ref{InterCurrGe}) then are opposite [see Eq.~(\ref{sp3functions})]. At the same time, the factors $d_{2k}^*d_{1i} = d_{2i}d_{1k}^*$ and $H_{qr}=H_{r'q'}$ are equal for these pairs. Therefore, the sum $\mathcal J_{qr}+\mathcal J_{r'q'}$, which contributes to the amplitude of the peak at $\Delta\mathbf R_{qr}$, is proportional to $(\mathbf j_{qr}-\mathbf j_{r'q'})\cdot\mathbf n = (\mathbf j_{qr}+\mathbf j_{q'r'})\cdot\mathbf n$.

In agreement with the previous discussion, we find computationally that the amplitude of the encircled peak in Fiq.~\ref{Fig_CurrFurGe111}a follows the sum $\sum_{qr}(\mathbf j_{qr}\cdot\mathbf n)+\sum_{q'r'}(\mathbf j_{q'r'}\cdot\mathbf n)$, where $\Delta\mathbf R_{qr}=-\Delta\mathbf R_{q'r'} = (a_0/4,a_0/4,a_0/4)$ and $C_q$ and $C_r$ are atoms that have index 1 and 2, $C_{q'}$ and $C_{r'}$ are atoms that have index 2 and 1, respectively (see Fig.~\ref{Fig_CurrFurGe111}b). The agreement between the time evolution of the amplitude of the peak and the sum of the currents is not as perfect as in the previous example, since they are both determined by several functions $\widetilde\phi_{1(2)i}$ per atom. The electron hole dynamics is faster than in the previous example, since the two outermost valence bands of germanium have a larger maximum energy splitting than in the outermost valence band of KBr (4.5 eV against 2.5 eV) (see Fig.~\ref{Fig_CurrFurKBr}b). The amplitude of the interatomic currents is larger than that in KBr, since spatial orbitals of Ge are delocalized to a much higher degree.
 
To illustrate the kind of information one would obtain with this method for nearest-neighbor atoms in Ge crystal, let some atom $C$ be situated at the center of our Ge cluster. We would be able to obtain the sum of interatomic electron hole currents from atom $C$ to its nearest neighbors minus the sum of the interatomic currents from the nearest neighbors to the next-nearest ones plus the sum of the currents from the second-nearest neighbors to the third-nearest ones and so on.

\begin{figure}[t]
\includegraphics{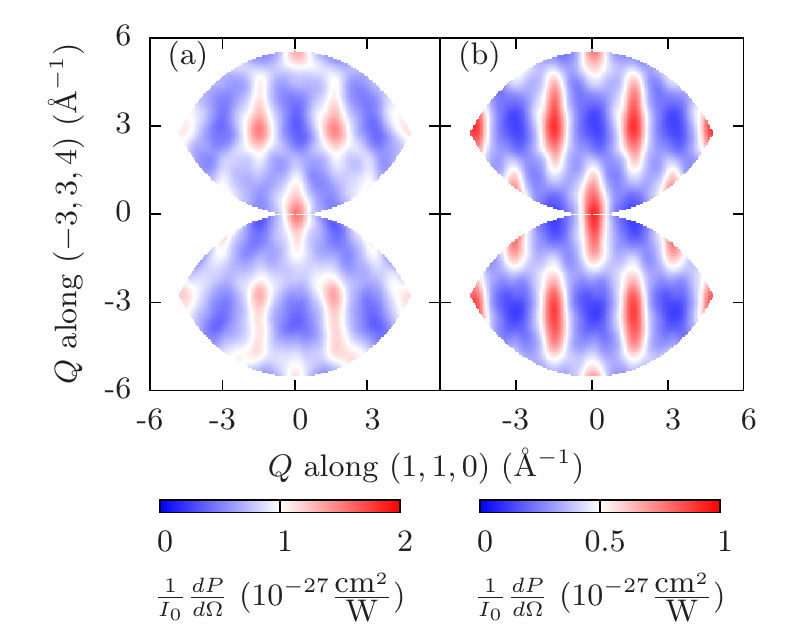}
  \caption{Scattering patterns at the probe-pulse intensity $I_0$ from the ionized Ge cluster obtained by the x-ray pulse polarized along $(1,1,0)$. $Q$ along direction perpendicular to the illustrated plane is zero. (a) No polarization filter is applied. (b) Polarization filter transmitting scattered photons polarized along $(1,1,0)$ is applied and the pattern is divided by the function $g(\mathbf Q)$. 
  }
\label{Fig_Gepatt110}
\end{figure}

\begin{figure}[t]
\includegraphics{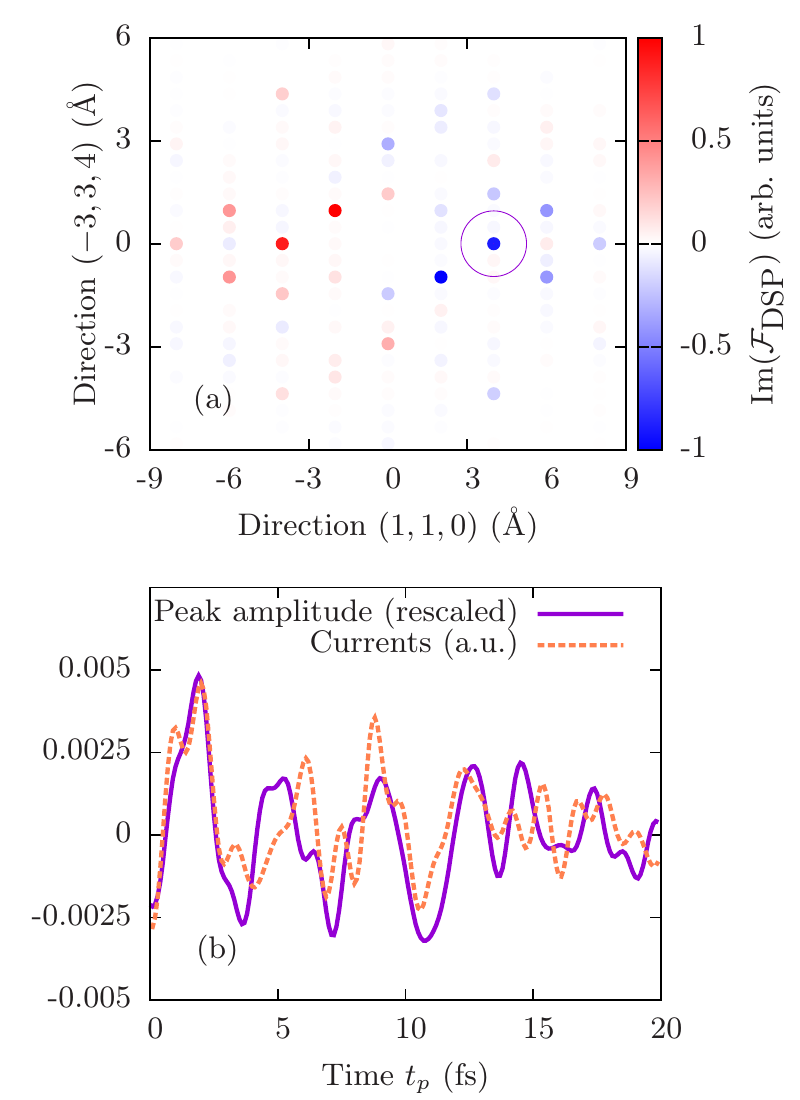}
  \caption{(a) The Fourier transform from $\mathbf Q$ space to real space of the scattering pattern in Fig.~\ref{Fig_Gepatt110}b. (b) Solid violet line: time evolution of the amplitude of the circled peak on panel (a). Orange dashed line: $\sum_{qr}|\mathbf j_{qr}|+\sum_{q'r'}|\mathbf j_{q'r'}|$, where $\Delta\mathbf R_{qr} = - \Delta\mathbf R_{q'r'} = (a_0/\sqrt{2},a_0/\sqrt{2},0)$, atoms $C_q$ and $C_r$ both having index 1, and atoms $C_{q'}$ and $C_{r'}$ both having index 2. }
\label{Fig_CurrFurGe110}
\end{figure}

The interatomic currents between second nearest-neighbor atoms in Ge crystal, which are separated by about 4 \r A, are also prominent in contrast to the case of KBr. We found that these currents have the largest component along the direction connecting the atoms, and their amplitudes can be found by applying a probe pulse polarized along the same direction. Figure \ref{Fig_Gepatt110} shows the scattering patterns with and without a filter for scattered photons obtained with a probe pulse with polarization $\boldsymbol\epsilon_\i=(\frac{1}{\sqrt2},\frac{1}{\sqrt{2}},0)$. The imaginary part of the Fourier transform of the pattern in Fig.~\ref{Fig_Gepatt110}b is shown in Fig.~\ref{Fig_CurrFurGe110}a. The encircled peak in Fig.~\ref{Fig_CurrFurGe110}a corresponds to atoms connected by the vector $(a_0/\sqrt{2},a_0/\sqrt{2},0)$, which is parallel to $\boldsymbol\epsilon_\i$. The plane of the scattering patterns depicted in Fig.~\ref{Fig_Gepatt110} was chosen such that this peak does not overlap with other prominent peaks. Similar to the case of nearest-neighbor atoms, the coefficient of proportionality between a factor $\mathcal J_{qr}(t_p)$ and $|\mathbf j_{qr}(t_p)|$ depends on the index of atoms $C_q$ and $C_r$, which now have the same indices in a pair. The time evolution of this peak approximately follows the time evolution of the sum $\sum_{qr}|\mathbf j_{qr}|+\sum_{q'r'}|\mathbf j_{q'r'}|$, where $\Delta\mathbf R_{qr} = - \Delta\mathbf R_{q'r'} = (a_0/\sqrt{2},a_0/\sqrt{2},0)$, atoms $C_q$ and $C_r$ both having index 1, and atoms $C_{q'}$ and $C_{r'}$ both having index 2 (see Fig.~\ref{Fig_CurrFurGe110}b). Note that although the projections of the currents between nearest-neighbor and next-nearest-neighbor atoms, depicted in Figs.~\ref{Fig_CurrFurGe111}b and \ref{Fig_CurrFurGe110}b, are of the same order, the amplitudes of the currents between nearest-neighbor atoms are approximately three times larger than the amplitudes of the currents between next-nearest-neighbor atoms.

It is also possible to find other polarizations $\boldsymbol\epsilon_\i$ of the probe pulse that will provide other projections $\mathbf j_{qr}\cdot\mathbf n$ of interatomic currents either between atoms connected by the vector $(a_0/4,a_0/4,a_0/4)$ or by other vectors. In order to determine the polarization $\boldsymbol\epsilon_\i$ of a probe pulse that provides a peak with an amplitude following some projection of the interatomic currents between atoms connected by a vector $\Delta\mathbf R_{qr}$, we performed an analysis of the integrals $d_{2k}^*d_{1i}$ and $j_{2k1i}=\int d^3r\widetilde\phi_{2k}^*(\mathbf r-\mathbf R_r)(\boldsymbol \nabla\cdot \mathbf n)\widetilde\phi_{1i}(\mathbf r-\mathbf R_q)$. It follows from Eq.~(\ref{dipintergrals}) that the integrals $d_{1i}$ and $d_{2k}$ do not depend on the sites on which the functions $\widetilde\phi_{1i}$ and $\widetilde\phi_{2k}$ are centered. The integral $j_{2k1i}$ also does not depend on the particular sites $C_q$ and $C_r$, but only on the vector $\Delta\mathbf R_{qr}$ between them. Therefore, one can compose two $4\times4$ matrices using the integrals $d_{1i}$, $d_{2k}$ and $j_{2k1i}(\Delta\mathbf R_{qr},\mathbf n)$ independently from specific atomic sites. The first matrix is a function of $\epsilon_\i^x$, $\epsilon_\i^y$ and $\epsilon_\i^z$, and its $ik$-th element is $d_{2k}^*d_{1i}$. The second matrix is a function of $n_x$, $n_y$ and $n_z$, and its $ik$-th element is $j_{2k1i}$ at a given vector $\Delta\mathbf R_{qr}$. Then, one has to find pairs of $\boldsymbol\epsilon_\i$ and $\mathbf n$ for which these two matrices are approximately proportional to each other. For analysis, we used the functions $\widetilde\phi_{1i}$ and $\widetilde\phi_{2k}$ from Eq.~(\ref{sp3functions}) to compose the matrices and determine the pairs of $\boldsymbol\epsilon_\i$ and $\mathbf n$. The resulting pairs of $\boldsymbol\epsilon_\i$ and $\mathbf n$ were the input for the numerical calculations of scattering patterns and interatomic currents. 
Since the matrix with the elements $d_{2k}^*d_{1i}$ is not a linear function of $\boldsymbol\epsilon_\i$, a superposition of such matrices at different polarizations can be a matrix that is not possible to obtain with a single $\boldsymbol\epsilon_\i$. Therefore, in principle, if it is necessary, one can measure several scattering patterns with different polarizations of the incoming beam and use a linear combination of their Fourier transforms in order to follow a certain projection of interatomic currents. 

\section{Conclusions and outlook}

In this paper, we have discussed the information that one can obtain, using the ultrafast RXS, about coherent nonperiodic electron dynamics in crystals. As an example of such dynamics, we considered electron hole motion in the valence bands of ionized crystals. The band width of the ultrashort probe x-ray pulse has to be larger than the width of the electron band where the dynamics takes place, in order to capture the dynamics. Since inelastic contributions unavoidably contribute to a scattering pattern obtained by such a pulse, ultrafast RXS from a nonstationary system provides in general information different from that of stationary RXS \cite{PopovaGorelova15}. Namely, we have shown that ultrafast scattering patterns are not centro-symmetric and do not resolve the standard structure factor. They still resolve structural information and, additionaly, can resolve the interatomic electron current between the scattering atoms. 

We have described a procedure to extract the interatomic electron currents from a single 2D scattering pattern by performing a Fourier transformation from $\mathbf Q$ space to real space. If a proper polarization of the incoming probe pulse has been chosen, the time evolution of the amplitude of a certain delta peak in the Fourier transform follows the time evolution of the sum of the interatomic currents between atoms connected by the same vector. The required polarization of the probe pulse can be determined by an analysis of the electron structure of the crystal, an example of which for Ge crystal is at the end of the last section. Although our analysis was performed for periodic crystals, our simulations on clusters reproduce these findings.

In this paper, we have considered just 2D scattering patterns. It should be possible to extract much richer information about the electron dynamics by measuring 3D scattering patterns of a sample. It seems likely that one could develop suitable phase-retrieval algorithms to obtain contributions by individual atomic pairs to a scattering pattern. In order to factor out the additional $\mathbf Q$ dependence in the scattering patterns due to different scattering angles at every $Q$ point, we have applied a polarization filter for scattered photons (see the discussion in Sec.~\ref{Subsec_ScatterinPatternKBr}). In principle, it should be possible to find a less experimentally sophisticated way to eliminate this additional $\mathbf Q$ dependence in the scattering patterns.

It follows from our estimate that both Ge and KBr crystals with one electron hole per $10^5$ Ge or Br atoms have to be irradiated with $\approx 10^{15}$ photons in order to achieve a signal of one photon per pixel on average. Subfemtosecond hard x-ray pulses that will be produced by free-electron lasers may contain $\approx 10^{10}$ photons \cite{ZholentsPRL04, EmmaPRL04, SaldinPhysRevSTAB06, KumarAppSciences13, TanakaPRL13,PratPRL15}. That means that the data can be accumulated using $10^5$ shots of such pulses, which can be produced in 20 minutes with a repetition rate of 100 Hz. As shown in Appendix \ref{App_RequiredPhotons}, the number of required photons is determined by the penetration depth of the x-ray beam, but does not depend on the interaction area. Therefore, the only restriction concerning the focal area of the probe x-ray beam is that it should be smaller than the pumped area of the sample and there is more freedom to lower the intensity of the beam in order not to damage a sample.

Since the goal of this study is to describe what information one can extract about the electronic motion independently from how it was launched, we have considered random electronic wave packets. Thus, it is demonstrated which information can be provided by ultrafast RXS in a general case of coherent electronic dynamics. When one studies scattering patterns in connection with some particular pump process, it should be possible to obtain even more comprehensive insight about the induced electronic motion.

We believe that our technique to study coherent electron dynamics in solids and to extract interatomic electron currents has a high potential in view of recent advances in creating coherent wave packets in crystals \cite{PolliNature07,KawakamiPRL10,KuehnPRL10,KuehnPRB10,SchubertNature14} and the growing role of ultrafast light-induced processes for high-speed electronics \cite{SchiffrinNature12,SchultzeNature12, HiroriNature11}, electro-optics \cite{GhimireNature11, SchubertNature14} and optical phase manipulations \cite{KawakamiPRL10,PolliNature07}.

\begin{appendix}
\section{Procedure to obtain a scattering pattern in a plane in $\mathbf Q$ space}
\label{App_Rotkin}
\begin{figure}[t]
\includegraphics{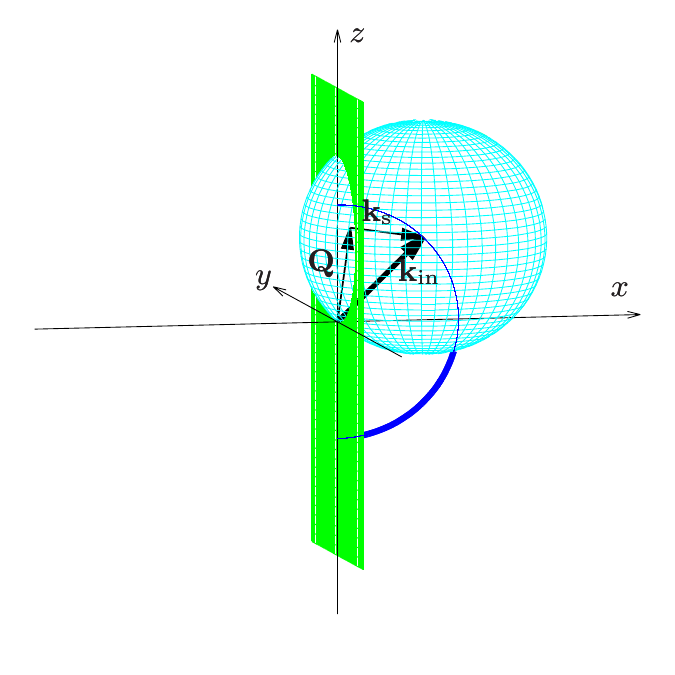}
  \caption{Rotation of the $\k_\i$ vector in the sample-fixed reference frame employed here. The sphere represents the Ewald's sphere associated with the depicted $\k_\i$ vector. The plane represents the plane with $Q$ points where the data is collected.}
\label{Fig_Rotkin}
\end{figure}

All scattering patterns in the article are presented in a $Q_{\shortparallel}Q_{\perp}$ plane at zero $Q_\dashv$, where $Q_{\shortparallel}$ is parallel to the polarization of the probe pulse, $\boldsymbol\epsilon_\i$, $Q_\perp$ is perpendicular to $\boldsymbol\epsilon_\i$, and $Q_\dashv$ is perpendicular to both $Q_{\shortparallel}$ and $Q_{\perp}$. The sample is at rest in this reference frame, but it has to be rotated in order to obtain data points in this plane. It is important for our study that $\boldsymbol\epsilon_\i$ is always parallel to a certain interatomic vector. In order to understand how the sample should be rotated in this case, let $Q_{\shortparallel}$ and $Q_{\perp}$ be aligned along the $y$ and $z$ directions, respectively. In the sample-fixed reference frame, we rotate $\mathbf k_\i$ instead of the sample. 

We rotate $\mathbf k_\i$ around the $y$ axis such that $\boldsymbol\epsilon_\i$ is always along the $y$ direction. Therefore, $\mathbf k_\i$ is rotated in such a way that it extends from the origin to a point that moves on the blue half circle as shown in Fig.~\ref{Fig_Rotkin}. The sphere in Fig.~\ref{Fig_Rotkin} is the Ewald's sphere centered at the depicted $\k_\i$ vector. The green plane in Fig.~\ref{Fig_Rotkin} is the $Q_{y} Q_{z}$ plane, where we aim to collect the data. The vector $\mathbf Q = \k_\i-\k_\s$ lies in the $Q_yQ_z$ plane if the vector $\mathbf k_\s$ points from some point on the plane to the center of the sphere. Since we focus on $\mathbf k_\s$ with $|\k_\s|\approx|\k_\i|$, $\mathbf k_\s$ must point from a point at the boundary of the sphere to its center. A vector $\mathbf k_\s$ satisfying both conditions is shown in Fig.~\ref{Fig_Rotkin}. Thus, $\k_\i$ provides $Q$ points on the $Q_yQ_z$ plane lying on the circle where this plane crosses the Ewald's sphere associated with $\k_\i$. The largest circle of $Q$ points is obtained for $\k_\i$ parallel to the $z$ axis, and just the zero point in the $Q_yQ_z$ plane is obtained for $\k_\i$ parallel to the $x$ axis.

The direction of $\k_\i$ that provides a given $Q$ point in the $Q_yQ_z$ plane, can be determined using three conditions: $|\k_\i|\approx |\k_\s|$, $\k_\i\perp\boldsymbol\epsilon_\i$ and $|\k_\i|=\omega_\i/c$. This gives us the following system of equations for $k_{\i x}$, $k_{\i y}$ and $k_{\i z}$:
\begin{align}
& (k_{\i x}-Q_x)^2 + (k_{\i y}-Q_y)^2 + (k_{\i z}-Q_z)^2=\frac{\omega_\i^2}{c^2},\nonumber\\
& k_{\i y} = 0,\\
& k_{\i x}^2+k_{\i y}^2+k_{\i z}^2 = \frac{\omega_\i^2}{c^2},\nonumber
\end{align}
with two solutions
\begin{align}
\mathbf k_\i = \lf(\pm \sqrt{\frac{\omega_\i^2}{c^2} -\frac{(Q_y^2+Q_z^2)^2}{4Q_z^2}},0,\frac{Q_y^2+Q_z^2}{2Q_z} \rt).\label{rotatedkin}
\end{align}
The solution with the positive $\k_{\i x}$ corresponds to vectors $\k_\i$ with the terminal point at the half circle in Fig.~\ref{Fig_Rotkin}. The square root in the expression for $\k_{\i x}$ limits the accessible Q points by two circles of radius $\omega_\i/c$ centered at points $(0, \pm\omega_\i/c)$ in the $Q_yQ_z$ plane. 

\section{Polarization filter for scattered photons}

\label{App_polarfilter}

In this Section, we derive the expression for the DSP under the assumption that the detector measures $I_{\k_\s}$, the intensity of light polarized along $\boldsymbol\epsilon_p$ and scattered with a scattering vector $\mathbf Q = \k_\i-\k_\s$. The derivation of the DSP in Ref.~\onlinecite{PopovaGorelova15} must be modified for the observable intensity \cite{Loudon}
\begin{align}
&\hat O_{\mathbf k_\s} = \frac{c}{2\pi}(\hat E^{-}(\mathbf r,t)\cdot\boldsymbol\epsilon_p^*)(\hat E^{+}(\mathbf r,t)\cdot \boldsymbol\epsilon_p),\\
&\hat E^{+}(\mathbf r,t) = i\sum_{s_\s} \sqrt{\frac{2\pi\omega_{\k_\s}}{V} }\boldsymbol\epsilon_\s\hat a_{\k_\s}e^{-i\omega_{\k_\s}t+i\k_\s\mathbf r},\\
&\hat E^{-}(\mathbf r,t) = -i\sum_{s_\s} \sqrt{\frac{2\pi\omega_{\k_\s}}{V} }\boldsymbol\epsilon_\s\hat a_{\k_\s}^\dagger e^{i\omega_{\k_\s}t-i\k_\s\mathbf r},
\end{align}
where the sum is over two possible polarization of $\k_\s$ and V is the quantization volume. Then,
\begin{align}
I_{\k_\s} = \lim_{t_f\rightarrow+\infty}\operatorname{Tr}[\hat \rho_f(t_f)\hat O_{\k_\s}],
\end{align}
where $\hat \rho_f(t_f)$ is the total density matrix of the electron system and the electromagnetic field at time $t_f$ after the action of the probe pulse. The resulting intensity does not depend on position $\mathbf r$ and time $t$. The expression for the DSP is then derived from $I_{\k_\s}$. 

The new condition of measurement results in substitution of polarizations $\boldsymbol \epsilon_{\s_{1,2}}$ in the terms $\sum_{s_\s}(\mathbf D_{J_{C_r}F}\cdot\boldsymbol \epsilon_\s)(\mathbf D_{FJ_{C_q}}\cdot\boldsymbol \epsilon_\s^*)$ in the expression for the DSP in Eq.~(\ref{DSP_DipApp_extended}) for their projections on $\boldsymbol\epsilon_p$. The projections of $\boldsymbol \epsilon_{\s_{1,2}}$ are $\boldsymbol\epsilon_p\cos(\nu_{1,2})$, where $\cos(\nu_{1,2})$ is the angle between $\boldsymbol \epsilon_{\s_{1,2}}$ and $\epsilon_p$: $ \cos(\nu_{1,2})=(\boldsymbol \epsilon_{\s_{1,2}}\cdot \boldsymbol\epsilon_p)$. Let $\boldsymbol\epsilon_{s_1}$ be perpendicular to $\boldsymbol\epsilon_p$, then $\boldsymbol\epsilon_{s_1} = [\k_\s\times\boldsymbol\epsilon_p]/|\k_\s|$ and $\boldsymbol\epsilon_{s_2} = [\k_\s\times[\k_\s\times\boldsymbol\epsilon_p]]/|\k_\s|^2$. Thus, the term $\sum_{s_\s}(\mathbf D_{J_{C_r}F}\cdot\boldsymbol \epsilon_\s)(\mathbf D_{FJ_{C_q}}\cdot\boldsymbol \epsilon_\s^*)$ in the new expression for the DSP turns into $g(\mathbf Q)(\mathbf D_{J_{C_r}F}\cdot\boldsymbol \epsilon_p)(\mathbf D_{FJ_{C_q}}\cdot\boldsymbol \epsilon_p^*)$, where the function
\begin{align}
&g(\mathbf Q) = ([\k_\s\times[\k_\s\times\boldsymbol\epsilon_p]]\cdot \boldsymbol\epsilon_p)^2c^4/\omega_\i^4 \label{PolarFilter}
\end{align}
depends on $\mathbf Q = \k_\i-\k_\s$ via Eq.~(\ref{rotatedkin}). We took into account that $|\k_\s|\approx|\k_\i|=\omega_\i/c$.

\section{Estimate of the required number of photons}
\label{App_RequiredPhotons}
The number of photons $N_{\text{ph}}$ that has to be sent on the sample in order to get a signal of one photon per pixel is given by $N_{\text{ph}} = (\la P_{\text{ph}}\ra N_{\text{h}})^{-1}$, where $ \la P_{\text{ph}}\ra$ is the average probability to scatter a photon into a pixel from a single scattering particle by a single incoming photon, and $N_{\text{h}}$ is the number of scattering particles, which are electron holes in our case. The number of the electron holes that interact with the probe pulse are given by $N_{\text{h}} = f_0 l_p d_{h}$, where $f_0$ is the interaction area, $l_p$ is the penetration depth of the x-ray beam and $d_{h}$ is the number of electron holes per unit volume. We assume that there is one electron hole per $10^5$ Br or Ge atoms in KBr or Ge crystals, respectively.

$ \la P_{\text{ph}}\ra$ is given by $f_0^{-1}\Omega_p \la d\sigma/d\Omega \ra $, where $\Omega_p$ is the pixel size and $\la d\sigma/d\Omega \ra$ is the mean differential scattering cross section. $\la d\sigma/d\Omega \ra$ can be derived from the mean DSP, which, as follows from Figs.~\ref{Fig_KBrpatt}, \ref{Fig_Gepatt111} and \ref{Fig_Gepatt110}, is approximately $I_0\times10^{-27}\text{cm}^2/\text{W}$ for a probe pulse with intensity $I_0$, duration $\tau_d = 200$ as and photon energy $\omega_\i$ on the order of 10 keV for both KBr and Ge crystals. Thus, $\la d\sigma/d\Omega \ra \approx 4\times 10^{-10}$ \r A$^2$ and 
\begin{align}
N_{\text{ph}} =  2.5 \times 10^{9}\text{ \r A}^{-2}/(l_p d_{h} \Omega_p  )\label{Nphotons}
\end{align}
 for both crystals.
%
%
The pixel size $\Omega_p$ is given by $(\lambda/\sqrt[3]{V_h})^2$, where $\lambda$ is the wavelength and $V_h = d_h^{-1}$ is the volume, where a single electron hole is distributed. 

We estimate the penetration depth in KBr and Ge crystals as $l_p = l_{m}/10$. $l_{m}$ is the lesser of the two mean free paths $(\sigma_{\text{res}}d_{h})^{-1}$ and $(\sigma_{\text{ion}}d_{cr})^{-1}$, where $\sigma_{\text{res}}$ is the total photoabsorption cross section of an ionized atomic bromine or germanium in the case of KBr or Ge crystals at the energy $\omega_{1s-4p}$ resonant with the $1s$ - $4p$ transition of Br or Ge, respectively. $\sigma_{\text{ion}} = \sigma_{\text{ion}}^{Br}+\sigma_{\text{ion}}^{K}$ is the sum of photoionization cross sections of neutral atomic bromine and potassium at the energy $\omega_{1s-4p}$ in the case of KBr crystal. Since the densities of neutral Br and K atoms, $d_{cr}^{Br}$ and $d_{cr}^{K}$, are equal in KBr, $\sigma_{\text{ion}}^Kd_{cr}^{K}+\sigma_{\text{ion}}^{Br}d_{cr}^{Br}= \sigma_{\text{ion}}d_{cr}$, where $d_{cr}=d_{cr}^{K}=d_{cr}^{Br} $. $\sigma_{\text{ion}}$ is the photoionization cross section of germanium at the energy $\omega_{1s-4p}$ and $d_{cr}$ is the atomic density of neutral Ge atoms in the case of Ge crystal. We calculate $\sigma_{\text{res}}$ and $\sigma_{\text{ion}}$ with the XATOM toolkit \cite{SonPRA11}.
 
There are four Br atoms in the cubic unit cell of KBr with the lattice parameter 6.6 \r A. Therefore, $d_{cr}=1.4\times 10^{-2}$ \r A$^{-3}$ and $d_h=1.4\times 10^{-7}$ \r A$^{-3}$. The wavelength of the incoming radiation $\lambda=0.9$ \r A results in a pixel size of $\Omega_p = 1\times 10^{-4}$ for KBr crystal. We obtain $\sigma_{\text{res}} = 2.5\times10^{-2}$ \r A$^2$, $\sigma^{Br}_{\text{ion}} = 2.9 \times 10^{-5}$ \r A$^2$ and $\sigma^{K}_{\text{ion}} = 2.3 \times 10^{-5}$ \r A$^2$ from the calculation, which results in $l_p = (\sigma_{\text{ion}}d_{cr})^{-1}/10 = 1.4 \times 10^5$ \r A. Thus, we obtain from Eq.~(\ref{Nphotons}) that $6\times10^{15}$ photons are needed for KBr in order to obtain a signal of one photon per pixel.
 
Ge crystal has 8 atoms in its cubic unit cell with a lattice parameter of 5.7 \r A. Therefore, $d_{cr}=0.04$ \r A$^{-3}$ and $d_h=4\times 10^{-7}$ \r A$^{-3}$. The wavelength of the incoming radiation $\lambda=1.1$ \r A results in a pixel size of $\Omega_p = 3\times10^{-4}$ for Ge crystal. We obtain $\sigma_{\text{res}} = 4.0\times10^{-2}$ \r A$^2$ and $\sigma_{\text{ion}} = 6.7 \times 10^{-5}$ \r A$^2$ from the calculation, which results in $l_p = (\sigma_{\text{ion}}d_{cr})^{-1}/10$ = $3\times10^4$ \r A. Thus, we find that $2\times10^{15}$ photons are required for Ge in order to obtain a signal of one photon per pixel.

The number of required photons may deviate in a real experiment. First, the estimate strongly depends on the density of holes, $N_{\text{ph}} \propto d_h^{-5/3}$. Second, if all electron holes move coherently in a crystal, a signal would be coherently enhanced. That means that one would probably see a pronounced Bragg pattern and fewer photons will be required to resolve a signal. However, if all electron holes behave differently, then the object size that has to be resolved would be larger than the subvolume there a single electron hole is situated. That means that the pixel size that we assumed would be smaller, and the number of required photons would be larger [see Eq.~(\ref{Nphotons})]. However, one may use the larger pixel size and average over the different dynamics of the various subvolumes.

\end{appendix}

\end{document}